\documentclass[preprint,12pt]{elsarticle}%
\usepackage{amssymb}
\usepackage{amsfonts}
\usepackage{amsmath}
\usepackage{graphicx}%
\providecommand{\U}[1]{\protect\rule{.1in}{.1in}}
\graphicspath{{E:/dottorato/Coulomb_damper/Testo/Paper/Immagini/}}
\journal{journal}

\begin{document}
%
\begin{frontmatter}%


%

\title{On roughness-induced adhesion enhancement}%

%

\author{M. Ciavarella}%
%

\address
{Politecnico di BARI. Center of Excellence in Computational Mechanics. Viale Gentile 182, 70126 Bari. Mciava@poliba.it}%
%

\begin{abstract}%

While adhesion reduction due to roughness is not surprising, roughness induced
adhesion remained a puzzle until recently Guduru and coworkers have shown a
very convincing mechanism to explain both the increase of strength and of
toughness in a sphere with concentric single scale of waviness. Kesari and
coworkers have later shown some very elegant convenient asymptotic expansion
of Guduru's solution. This enhancement is very high and indeed, using Kesari's
solution, it is here shown to depend \textit{uniquely} on a Johnson parameter
for adhesion of a sinusoidal contact. However, \textit{counterintuitively}, it
leads to unbounded enhancement for conditions of large roughness for which
Johnson parameter is very low. Guduru postulated that this enhancement should
occur after sufficiently large pressure has been applied to any spherical
contact. Also, that although the enhancement is limited to the JKR regime of
large soft materials with high adhesion, the DMT limit for the smooth sphere
is found otherwise. However, for hard materials, even the DMT limit for the
smooth solids is very hard to observe, which suggest that also adhesion
reduction is yet not well understood.

The limitations of the assumption of simply connected area is here further
discussed, and a well known model for hard particles in contact with rough
planes due to\ Rumpf is used to show that in the range where unbounded
increase is predicted, orders of magnitude reduction is instead expected for
rigid solids. We suggest that Guduru's model may be close to an upper bound to
adhesion of rough bodies, while the Rumpf-Rabinovich model may be close to a
lower bound.%

\end{abstract}%
%

\begin{keyword}%

Roughness, Adhesion, Guduru's theory, Fuller and Tabor's theory%

\end{keyword}%
%

\end{frontmatter}%



\section{\bigskip Introduction}

Fuller and Tabor (1975) were perhaps the first to measure adhesion of low
modulus materials like smooth rubber lenses against roughened surfaces, and
clearly showed very small amounts of roughness amplitude (few microns) were
sufficient to destroy adhesion almost completely. They then proceeded to
develop an asperity model, which is commonly believed not to permit
enhancement of adhesion, but instead very rapid extinction\footnote{In
reality, reduction is \textit{assumed} from the outset in the model, as Fuller
and Tabor postulated that the smooth sphere case corresponded to the aligned
asperity case in the nominally plane model.}. In nature, various mechanisms
have been shown to lead to adhesion enhancements in insects which use adhesion
for their locomotion including varying the shape of each contact as well as
increasing the number of them, so as to obtain a benefit from contact
splitting (Hui \textit{et al}., 2004, Kamperman \textit{et al}, 2010, Gao
\&\ Yao, 2004, Yao \&\ Gao, 2006). Adhesion enhancement was measured by Briggs
\& Briscoe (1977), and Fuller \& Roberts (1981), who were unable to explain
the data, particularly the increase of energy dissipation. Persson (2002)
postulated that an increase of adhesion may occur for the increase of surface
area induced by roughness, but later on this has been shown to be not the
reason for the enhancement clearly explained by Guduru and collaborators
(Guduru (2007), Guduru \&\ Bull (2007), Waters \textit{et al} (2009)).
Guduru's solution is a very classical contact mechanics exact solution
assuming a simply connected contact area develops in a spherical contact
having a concentric axisymmetric waviness. Clearly, as it was apparent also to
Guduru (2007), there are some limitations for this solution to hold, as one
expects the contact to occur only on the crests of the waviness, for
"sufficiently" large amplitude roughness and this would limit the enhancement
shown in the simply connected contact area model. The separated contact
solution is unfortunately not possible in closed form, and therefore it was
not studied by Guduru, who noticed however the is a large set of conditions
for which we could assume it holds. Guduru's theory is very brilliantly
described and corroborated by experiments, and serves the main purpose of
showing when the enhancement can occur. However, we move here from a opposite
perspective, trying to understand why the large enhancement is not commonly
observed. We will therefore discuss two main aspects of the solution: the
assumption of simply connected contact area, and the assumption of JKR regime.

Regarding the first aspect, Guduru (2007) writes a condition on the
monotonicity of the profile function which guarantees a simply connected area,
but limits the amplitude of roughness in fact to regions where the enhancement
is not too large. Guduru recognized that the condition is over-restrictive, in
that, for sufficiently large pressure during loading, we may expect the simply
connected area to being established, and therefore full enhancement may occur.
Regarding the second aspect, Waters \textit{et al} (2009) develop a
Maugis-Dugdale solution with an annulus of uniform tension at the edge of the
contact area, showing enhancement is limited to the JKR regime. However, they
seem to imply, at least with the limited set of parameters they study, the
limit should be that of a DMT smooth sphere, which in the case of a rough
sphere, is in fact, although not an enhancement, still very high. We will
therefore consider a limit case, that of rigid bodies in contact, showing that
the limit is more complex than that, and permitting a very large reduction
indeed. Much of the debate on adhesion of spherical bodies has concentrated on
the transition from the DMT theory (see Greewood, 2007 for a detailed
treatment about this controversial theory) to the JKR limit. However, the two
theories differ, for pull-off, only on a small prefactor (3/2 in JKR, and 2 in
DMT), whereas the more we understand about the effect of roughness on
adhesion, the more we are confused between orders of magnitude reduction (as
it is commonly observed) and 1 order of magnitude enhancement (or higher?), as
Guduru's theory shows. This note therefore attempts to compare various results
in order to hopefully arrive at a better comprehension between these extreme limits.

In some of the developments, we use the recently developed asymptotic
expansion of Guduru's solution developed by Kesari \textit{et al} (2010,2011),
finding some reduced parameter dependence in the solution, which in fact is
exactly a JKR solution, different in the loading and the unloading phase. We
examine therefore where this solution holds, therefore providing some
hopefully insight in further understanding of the more general complex problem
of adhesion in rough surfaces which, in view of the competition between
enhancement and reduction mechanisms, is still not well understood.

\section{Guduru-Kesari theory}

\bigskip Waters \textit{et al} (2009) give a good summary of Guduru's theory
and experiments. They have a surface defined as $f\left(  r\right)
=\frac{r^{2}}{2R}+A\left(  1-\cos\frac{2\pi r}{\lambda}\right)  $, where $R$
is the sphere radius, $\lambda$ is wavelength of roughness, $h$ is its
amplitude, and $A$ can be both positive in the case of a central convex
asperity, and negative, for a central concave trough. The solution of the
contact problem with adhesion is possible assuming a simply connected contact
area, as a function of two parameters
\begin{equation}
\alpha_{G}=\frac{AR}{\lambda^{2}}\quad,\quad\beta_{G}=\frac{\lambda^{3}%
E^{\ast}}{2\pi wR^{2}}%
\end{equation}
where $w$ is the surface energy, and $E^{\ast}$ the plane strain elastic
modulus. The results are given in terms of a dimensionless load $\widehat
{P}=P/\left(  \frac{3}{2}\pi wR\right)  $ so that $\widehat{P}=1$ corresponds
to the smooth sphere, and also for dimensionless approach $\widehat
{h}=hR/\lambda^{2}$, and contact radius $\widehat{a}=a/\lambda$, as%
\begin{align}
\widehat{P}_{G}\left(  \widehat{a}\right)   &  =\frac{8}{3}\beta\left[
\frac{2}{3}\widehat{a}^{3}+\alpha\left(  \frac{4\pi^{2}\widehat{a}^{3}}%
{3}+\frac{\pi\widehat{a}}{2}H_{1}\left(  2\pi\widehat{a}\right)  -\pi
^{2}\widehat{a}^{2}H_{2}\left(  2\pi\widehat{a}\right)  \right)  \right]
-\frac{8}{3}\sqrt{\beta\widehat{a}^{3}}\\
\widehat{h}_{G}\left(  \widehat{a}\right)   &  =\widehat{a}^{2}+\alpha\pi
^{2}\widehat{a}H_{0}\left(  2\pi\widehat{a}\right)  -\sqrt{\widehat{a}/\beta}%
\end{align}

Kesari \textit{et al} (2010,2011) have developed a very elegant "envelope"
solution of the Guduru problem. The envelop is obtained by joining in an
asymptotic solution for small roughness sizes, in particular $\lambda<<a$:
said otherwise, there much be enough wavelengths of roughness in the contact
area --- a condition quite close to what was used in the Archard cascade
process of redistribution of loads in the adhesionless contact problem
(Ciavarella \textit{et al.} (2000)), and which can be checked a posteriori.
They notice that the solution has some simplified behaviour but do not further
discuss this aspect. The envelop solution in terms of load and indentation
depth is%
\begin{align}
P_{K}\left(  a\right)   &  =\frac{4}{3R}E^{\ast}a^{3}-a^{3/2}\left(
\sqrt{8\pi wE^{\ast}}\pm2\pi E^{\ast}\frac{A}{\sqrt{\lambda}}\right)
\label{kesari_load}\\
h_{K}\left(  a\right)   &  =\frac{a^{2}}{R}-a^{1/2}\left(  \sqrt{\frac{2\pi
w}{E^{\ast}}}\pm\pi\frac{A}{\sqrt{\lambda}}\right)  \label{kesari_approach}%
\end{align}
where we have grouped the term with the same power in contact radius $a$ ---
the theory obviously corresponds to the known JKR theory for $\frac{A}%
{\sqrt{\lambda}}=0$.

\subsection{Reformulation of Kesari equations}

An obvious remark about the equations (\ref{kesari_load},\ref{kesari_approach}%
) is that they are \textit{exactly} those of the JKR theory also in the case
of roughness, but with a corrected (\textit{enhanced} or \textit{reduced},
respectively for unloading or loading) surface energy
\begin{align}
P\left(  a\right)   &  =\frac{4}{3R}E^{\ast}a^{3}-a^{3/2}\sqrt{8\pi wE^{\ast}%
}\left(  1\pm\frac{1}{\sqrt{\pi}\alpha_{KLJ}}\right) \\
h\left(  a\right)   &  =\frac{a^{2}}{R}-a^{1/2}\sqrt{\frac{2\pi w}{E^{\ast}}%
}\left(  1\pm\frac{1}{\sqrt{\pi}\alpha_{KLJ}}\right)
\end{align}
where
\begin{equation}
\alpha_{KLJ}=\sqrt{\frac{2w\lambda}{\pi^{2}E^{\ast}A^{2}}}%
\end{equation}
is the parameter Johnson (1995) introduced for the JKR adhesion of a nominally
flat contact having a single scale sinusoidal waviness of amplitude $A$ and
wavelength $\lambda$. We can also recast the JKR "envelope" equations of
Kesari \textit{et al} (2011)\ in terms of only one of the Guduru dimensionless
parameters, $\beta_{G}$ together with $\alpha_{KLJ}$%
\begin{align}
\widehat{P}_{K}\left(  \widehat{a}\right)   &  =\frac{8}{3}\beta_{G}%
\widehat{a}^{3}-4\sqrt{\widehat{a}^{3}\beta_{G}}\left(  1\pm\frac{1}{\sqrt
{\pi}\alpha_{KLJ}}\right) \label{kesari1}\\
\widehat{h}_{K}\left(  \widehat{a}\right)   &  =\widehat{a}^{2}-\left(
\frac{\widehat{a}}{\beta_{G}}\right)  ^{1/2}\left(  1\pm\frac{1}{\sqrt{\pi
}\alpha_{KLJ}}\right)  \label{kesari2}%
\end{align}
\bigskip

It is therefore very interesting that the increase of adhesion should scale
with the same parameter of sinusoidal contact. It is also very interesting and
unexpected a priori that an important value for this parameter in eqt.
(\ref{kesari1}, \ref{kesari2}) is
\begin{equation}
\alpha_{KLJ}=\frac{1}{\sqrt{\pi}}\simeq0.56
\end{equation}
which is exactly the value for which the sinusoid self-flattens to full
contact under no applied load. In fact, the behaviour of a sinusoidal contact
should be explained in few words, following Fig.1, where $\alpha_{KLJ}$ is
indicated as $\alpha_{0}$ because in general, for a multiscale roughness, this
parameter can only be defined appropriately to a single sinusoid, and in
particular tends to increase for low fractal dimensions, which is the common
case observed in practice (see Afferrante \textit{et al,} 2015).

\begin{center}
$%
\begin{array}
[c]{cc}%
{\includegraphics[
height=4.1226in,
width=6.3097in
]%
{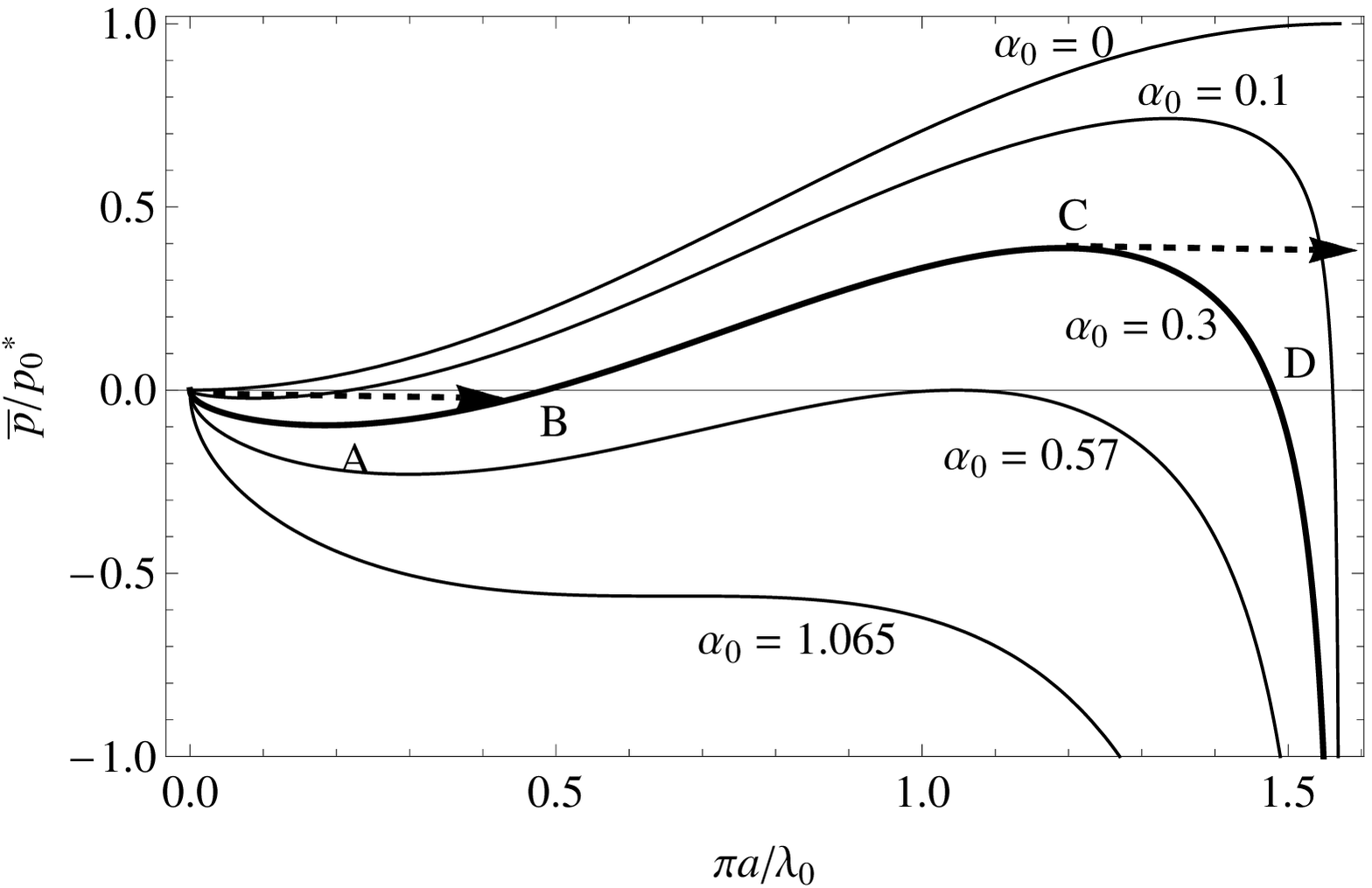}%
}
&
\end{array}
$\bigskip

Fig.1. Behaviour of a nominally flat single sinusoidal contact
\end{center}

Suppose we follow the curve corresponding to $\alpha_{0}=0.3$. Under zero
load, since the curve is decreasing (and hence unstable), the contact will
"jump" to point B, similarly to any JKR solution. From this point on, a
compressive load $\overline{p}$, moves the system along the curve BC, until at
point C, again instability occurs and complete contact occurs. The maximum
mean pressure needed to establish full contact is a fraction $\kappa$ of
$p_{0}^{\ast}$, where $p_{0}^{\ast}=\pi E^{\ast}A/\lambda$ is the pressure for
full contact without adhesion, and reduces with $\alpha_{0}$. For $\alpha
_{0}>\alpha_{cr}\simeq0.56$, partial contact does not occur and the surfaces
immediately snap together until full contact occurs. At that point, when
contact is established, it can be maintained also for negative arbitrarily
high (tensile) mean pressures, provided $\bar{p}\geq\eta p_{0}^{\ast}$, where
$\eta$ is a negative parameter, function of $\alpha_{0}$.

Notice there is a discussion in Waters \textit{et al} (2009), based on
qualitative energy balance, which essentially repeats the same process in
Johnson's sinusoid $\alpha_{KLJ}>0.56$, which for Guduru's parameter becomes
gives
\begin{equation}
\alpha_{G}<\frac{1}{\sqrt{2\pi\beta_{G}}}%
\end{equation}
Waters \textit{et al} (2009) remark, correctly, that this condition only
\textit{approximately} estimates for which parameters the roughness is
flattened and simply connected area is established spontaneously --- that is,
within the macroscopic contact area.

Returning to the actual Guduru model of a spherical contact with sinusoidal
waviness, it wasn't necessary that the same parameter relative to the waviness
in Guduru's solution appeared, and indeed, \textit{it did not appear} in
Guduru's solution, but only in its envelope from Kesari's solution. Assuming a
simply connected area, the condition $\alpha_{KLJ}>0.56$ seems sufficient to
guarantee that this solution is appropriate, but as we shall see, enhancement
is this region is relatively low, and instead, surprisingly, the largest
enhancement occurs for low $\alpha_{KLJ}$. We can estimate the actual "radius
of spontaneous contact" therefore precisely from the full Guduru solution, and
indeed also approximately from Kesari's solution.

It would seem that $\alpha_{KLJ}>0.56$ shows an important transition too: the
loading curve envelope becomes Hertzian for this value, and "less than
Hertzian" for lower values $\alpha_{KLJ}<0.56$, which in contrast would be the
values where the enhancement of pull-off load would be even more than 4. In
particular, an effective energy on loading and unloading, respectively, can be
defined as
\begin{equation}
\sqrt{w_{eff,loading}}=w\left(  1-\frac{1}{\sqrt{\pi}\alpha_{KLJ}}\right)
\quad;\quad w_{eff,unloading}=w\left(  1+\frac{1}{\sqrt{\pi}\alpha_{KLJ}%
}\right)  ^{2}%
\end{equation}
where we have left the square root for $w_{eff,loading}$ to take into account
that for $\alpha_{KLJ}<1/\sqrt{\pi}$, $\sqrt{w_{eff,loading}}$ should be negative.

Notice that for the Kesari expansion to be valid, one needs $\widehat{a}>>1.$
Therefore, for the condition at pull-off to be reasonably evaluated from this
analysis, one needs in general \textit{low values of }$\beta_{G}<<1$. In this
limit only, one could use pull-off from the unloading curve, so the size of
contact area at pull-off is
\begin{equation}
\widehat{a}_{c,low\beta_{G}}^{3}=\frac{9\pi}{4}\frac{R^{2}}{\lambda^{3}}%
\frac{w}{E^{\ast}}\left(  1+\frac{1}{\sqrt{\pi}\alpha_{KLJ}}\right)
^{2}=\frac{9}{8\beta_{G}}\left(  1+\frac{1}{\sqrt{\pi}\alpha_{KLJ}}\right)
^{2}%
\end{equation}
and the actual value of pull-off is%

\begin{equation}
\widehat{P}_{\min,low\beta_{G}}=-\left(  1+\frac{1}{\sqrt{\pi}\alpha_{KLJ}%
}\right)  ^{2} \label{pmin}%
\end{equation}

Instead, the size of the contact area when the load is zero, is obtained from
the \textit{loading} curve (only for $\alpha_{KLJ}>1/\sqrt{\pi}$ ) as%
\begin{equation}
\widehat{a}_{eq,low\beta_{G}}^{3}=\frac{9}{2\beta_{G}}\left(  1-\frac{1}%
{\sqrt{\pi}\alpha_{KLJ}}\right)  ^{2}%
\end{equation}
and this suggests an alternative map from Fig.5 of Waters \textit{et al}
(2009) where dependence on $\alpha_{G},\beta_{G}$ is shown. Indeed, in view of
the convenience of writing Kesari's equation in terms of $\alpha_{KLJ}$,
perhaps a clearer notation is to rewrite the Guduru equations in terms of
$\beta_{G},\alpha_{KLJ}$ instead of their original $\alpha_{G},\beta_{G}$
using $\alpha_{G}=\frac{1}{\pi^{3/2}\alpha_{KLJ}\sqrt{\beta_{G}}}$ and
obtaining
\begin{align}
\widehat{P}_{G}\left(  \widehat{a}\right)   &  =\frac{8}{3}\beta_{G}%
\widehat{a}^{3}+\left[  \frac{4\sqrt{\beta_{G}}}{\pi^{3/2}\alpha_{KLJ}}\left(
\frac{4\pi^{2}\widehat{a}^{3}}{3}+\frac{\pi\widehat{a}}{2}H_{1}\left(
2\pi\widehat{a}\right)  -\pi^{2}\widehat{a}^{2}H_{2}\left(  2\pi\widehat
{a}\right)  \right)  \right]  -4\sqrt{\beta_{G}\widehat{a}^{3}}\\
\widehat{h}_{G}\left(  \widehat{a}\right)   &  =\widehat{a}^{2}+\left[
\frac{\pi^{1/2}\widehat{a}H_{0}\left(  2\pi\widehat{a}\right)  }{\alpha
_{KLJ}\sqrt{\beta_{G}}}\right]  -\sqrt{\frac{\widehat{a}}{\beta_{G}}}%
\end{align}
where the terms under square parentheses cause the fluctuation in Guduru's
equations, and are substituted in the envelop by the factors $\left(
1\pm\frac{\sqrt{\pi}}{\alpha_{KLJ}}\right)  .$

In fig.2,3 we show some examples of load-approach, approach-area and
load-area, in 2 interesting cases with low $\beta_{G}$: where Kesari's
envelope (represented also here as blue and red curves) works well, and with
low and with high $\alpha_{KLJ}$. At low $\alpha_{KLJ}$, the enhancement in
pull-off is greater and the loading curves actually fold on each other in the
load-approach diagram (Fig.1a). As the Kesari's equation predict, the
spontaneous jump into contact practically does not exist, and therefore the
large enhancement can only be obtained with sufficiently large pressure during
the loading stage, and Kesari \textit{et al} (2010,2011) give also additional
results on how to compute the loading and unloading curves, as well as the
integral of the envelop curve to compute energy dissipation.

\begin{center}
$%
\begin{array}
[c]{cc}%
{\includegraphics[
height=2.4526in,
width=3.659in
]%
{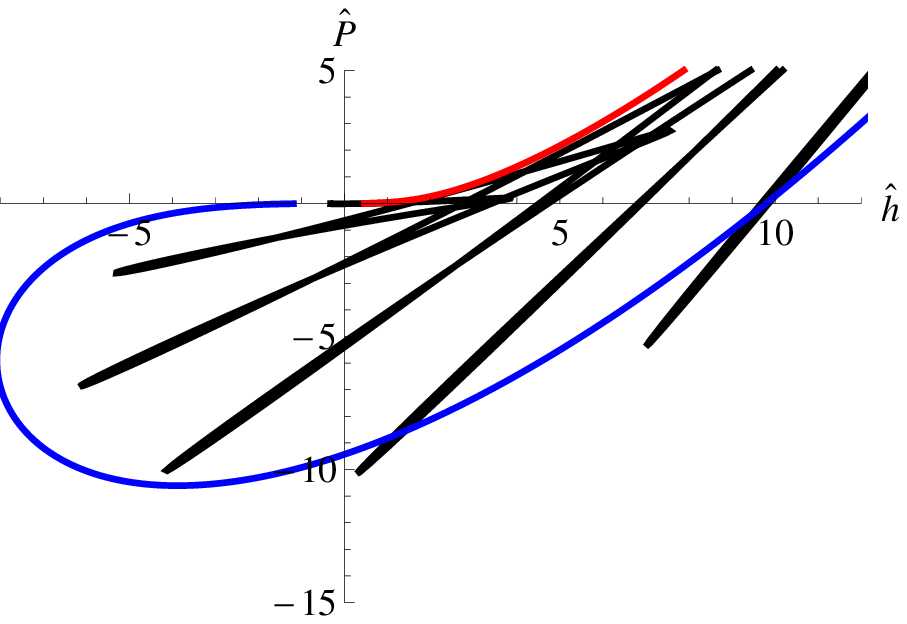}%
}
& (a)\\%
{\includegraphics[
height=2.3687in,
width=3.659in
]%
{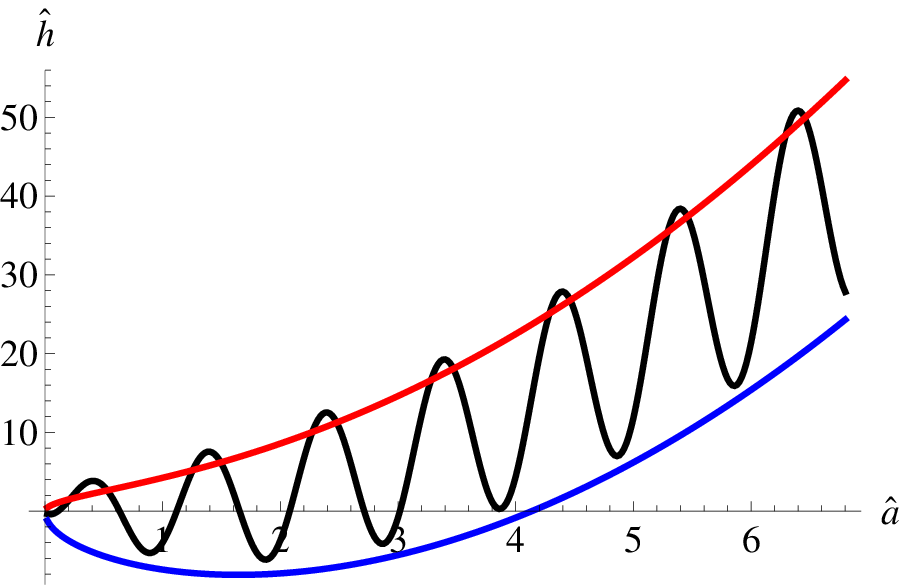}%
}
& (b)\\%
{\includegraphics[
height=2.3255in,
width=3.659in
]%
{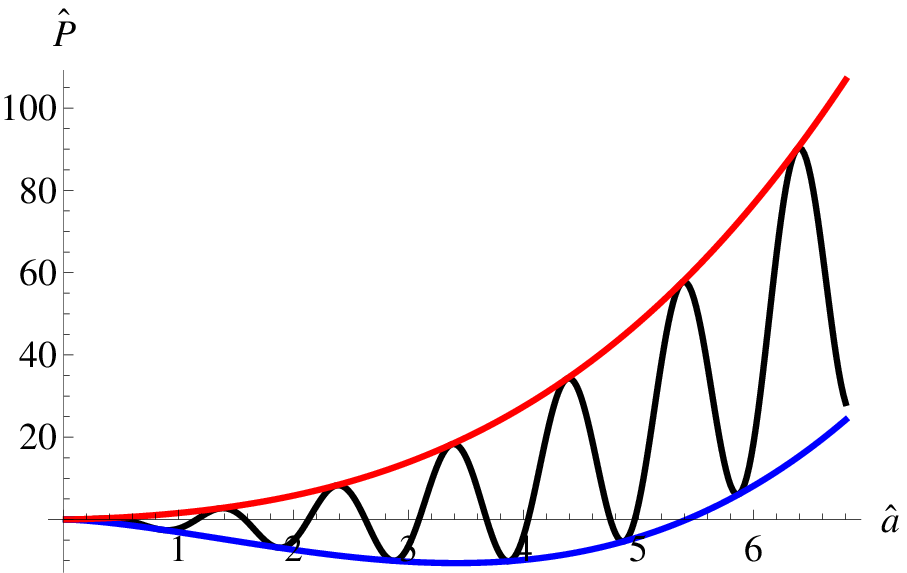}%
}
& (c)
\end{array}
$

Fig.2 - Load-approach (a), approach-area (b) and load-area (c), for low
$\alpha_{KLJ}=0.25$ and low $\beta_{G}=0.15$. Blue and red lines are the
Kesari envelopes for unloading and loading, respectively. Notice that in this
case of $\alpha_{KLJ}=0.25$ the curves on loading fold on each other.

$%
\begin{array}
[c]{cc}%
{\includegraphics[
height=2.4396in,
width=3.659in
]%
{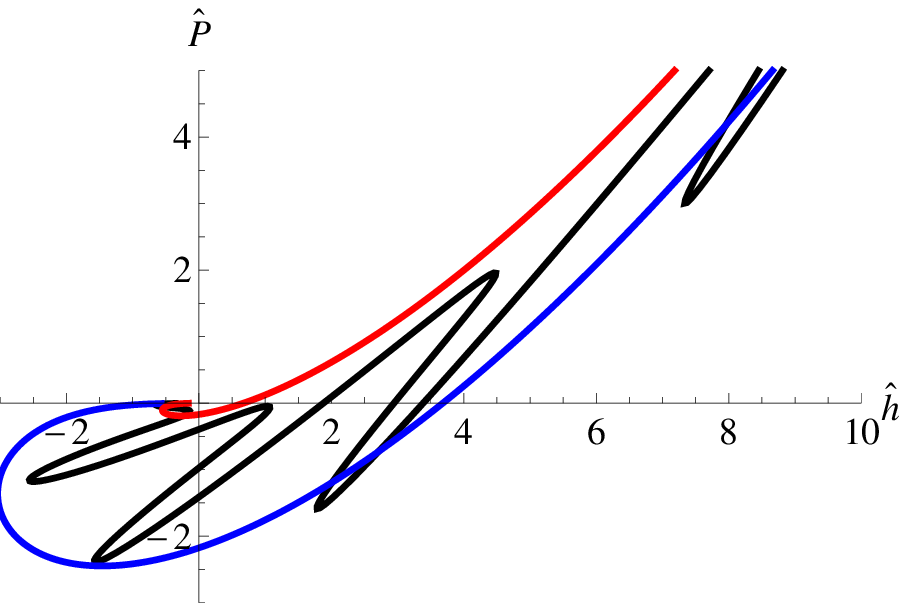}%
}
& (a)\\%
{\includegraphics[
height=2.3964in,
width=3.659in
]%
{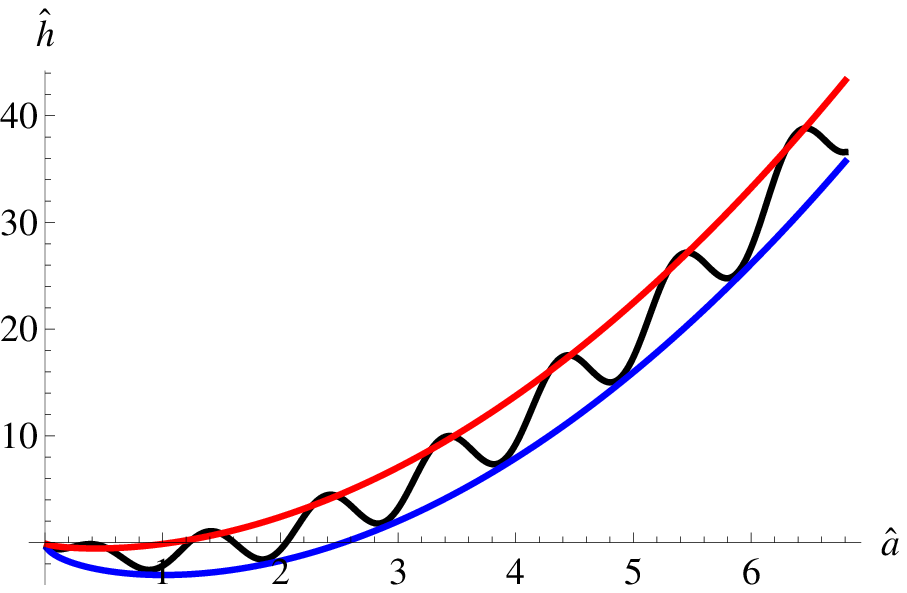}%
}
& (b)\\%
{\includegraphics[
height=2.4396in,
width=3.659in
]%
{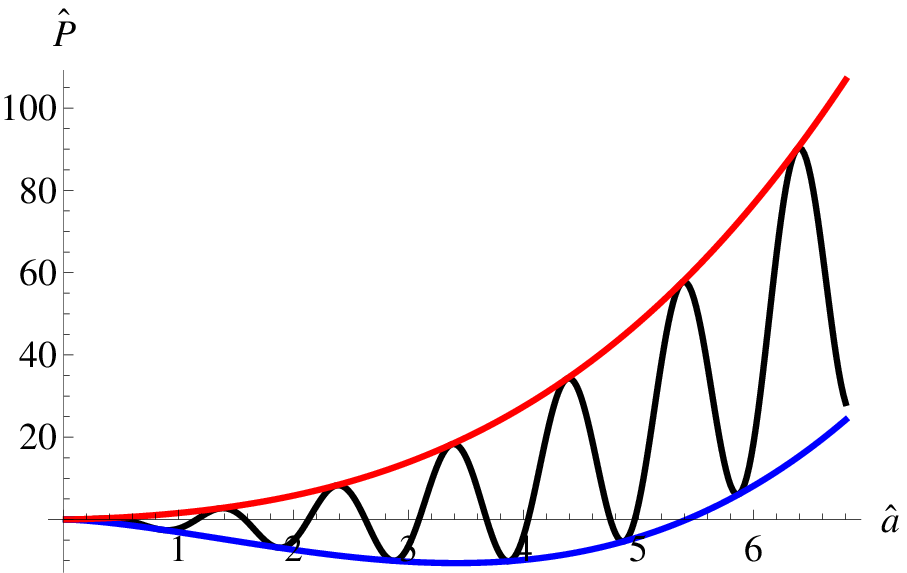}%
}
& (c)
\end{array}
$

Fig.3 - Load-approach (a), approach-area (b) and load-area (c), for high
$\alpha_{KLJ}=1$ and low $\beta_{G}=0.15$.Blue and red lines are the Kesari
envelopes for unloading and loading, respectively
\end{center}

For larger $\alpha_{KLJ},$ instead, we have more noticeable spontaneous jump
into contact and in this particular case, unloading from this point already
seems to lead to a value of pull-off close to the Kesari envelope. Notice that
Guduru (2007) had attempted an empirical fit for the pull-off enhancement of
the type
\begin{equation}
\widehat{P}_{\min}=-\left(  1+\frac{C}{w^{\beta}}\right)
\end{equation}
where $\beta$ was observed to vary between 0.5 and 0.9. The equation above
(\ref{pmin}) seems to justify this, considering that\ $\alpha_{KLJ}\sim
\sqrt{w}$, but does not permit a better fit including the deviations from
Kesari's expansion at large $\beta_{G}$.

The enhancement of pull-off defined as $\left\vert \widehat{P}_{\min
}\right\vert $ is shown in Fig.4 in the asymptotic limit at low $\beta_{G}$
i.e. from eqt.(\ref{pmin}). It is clear that enhancements are small when
$\alpha_{KLJ}>10$ (about 10\% increase, and less), significant in
$\alpha_{KLJ}<5$, has a value of 4 at the critical value $\alpha_{KLJ}=0.56$
and continues to increase with lower $\alpha_{KLJ}.$ This is significant as
the factor $\alpha_{KLJ}$ measures the how sticky is the waviness in itself,
and all the equations are saying is that a very sticky small amplitude
waviness doesn't add nor destroy the stickiness of the macroscopic sphere. The
biggest enhancement would seem to occur, surprisingly perhaps, at low values
of $\alpha_{KLJ}$, when the value of pull-off in dimensional terms would seem
to be
\begin{equation}
P_{\min,\lim}\rightarrow-\frac{3}{2}\pi wR\left(  \frac{1}{\sqrt{\pi}%
\alpha_{KLJ}}\right)  ^{2}=-\frac{3}{4}\pi^{2}E^{\ast}R\left(  \frac{A}%
{\sqrt{\lambda}}\right)  ^{2}%
\end{equation}
which no longer depends on surface energy and seems instead to be related to
elastic modulus alone and geometrical quantities\footnote{Let us check this
equation for example with atomic roughness, of both amplitude and wavelength
$a_{0}$. As $\frac{w}{E^{\ast}}=l_{a}$, and for Lennard-Jones potential,
$l_{a}/a_{0}=0.05$%
\begin{equation}
P_{\min,\lim}\rightarrow-\frac{3}{4}\pi^{2}E^{\ast}Ra_{0}=-15\pi^{2}wR=-148wR
\end{equation}
which in fact corresponds to an enhancement with respect to the smooth sphere
(?) of a factor $\frac{148}{1.5}=100$. But one could take wavelength of size
$a_{0}$ and higher amplitude, and this would grow even further with $A^{2}$
without any apparent limit.}. This limit does not have much sense but it is
interesting to remark that, while it is easy to understand why the uniform
pressure shape of a contact could lead to theoretical strength in the contact
(Gao \& Yao, 2004), although this limit is sensitive to shape at large sizes,
here we seem to reach very high adhesion limits with a spherical contact, of
very large size, adding roughness, which normally is instead considered to be
a detrimental factor.

\bigskip Clearly, there must be a limit when roughness separates the contact
into multiply connected areas: in that region, pull-off enhancement may still
be possible if a contact splitting effect prevails over the effect of
competing elastic deformation of the asperities at different heights as
distributed over the spherical shape.

\begin{center}
$%
\begin{array}
[c]{cc}%
{\includegraphics[
height=2.3699in,
width=3.6611in
]%
{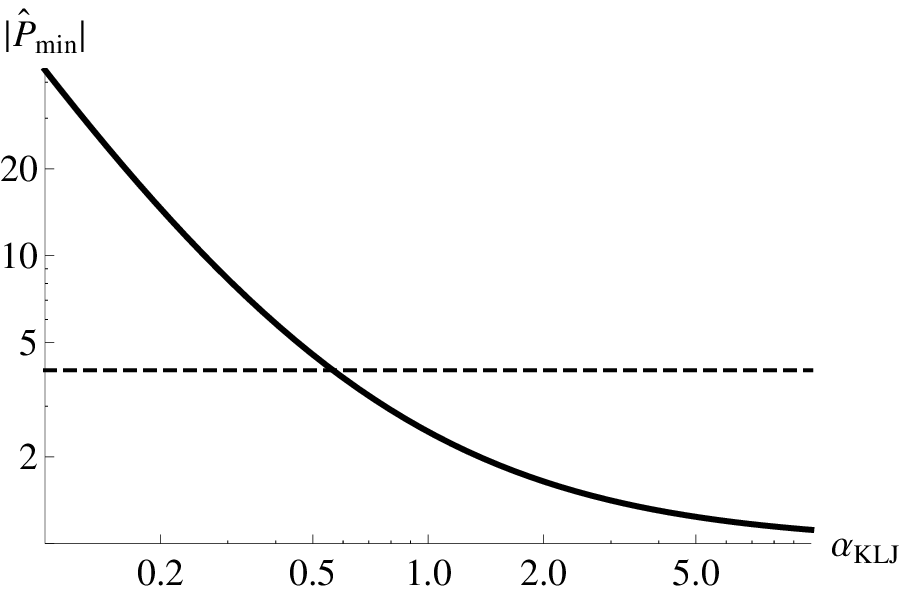}%
}
&
\end{array}
$

Fig.4 - Asymptotic dependence of pull-off enhancement on $\alpha_{KLJ}%
=\sqrt{\frac{2w\lambda}{\pi^{2}E^{\ast}A^{2}}}$ for \textit{low values of
}$\beta_{G}<<1$\ . Dashed line is the value of enhancement (4) for
$\alpha_{KLJ}=0.56$
\end{center}

\section{\bigskip Limits on the enhancement}

There are various reasons why unbounded enhancement doesn't occur. Even those
aspects which were already discussed by Guduru and coworkers are given here in
more details, in the new notation permitted by Kesari's equation, and new
comparisons are added.

\subsection{Deviations at large $\beta_{G}$}

As we have discussed in the previous paragraph, the Kesari expansion is valid
in the limit $\widehat{a}>>1$, which implies in general \textit{low values of
}$\beta_{G}<<1.$ As clear in the Fig.5 which compare the actual pull-off
values from the full Guduru model with those of the asymptotic expansion, the
enhancement is much reduced at large $\beta_{G}$. In particular, following the
various curves in Fig.5 at increasing $\beta_{G}$, the enhancement is much
reduced for intermediate values of $\alpha_{KLJ}$ although the low
$\alpha_{KLJ}$ paradoxical behaviour seems preserved. Notice we can actually
switch to reduction instead of enhancement for large $\beta_{G}$, but this is
due, as noticed by Guduru (2007), to the fact that in the limit of very large
wavelength (which is also large $\beta_{G}$), we have essentially only one
contact at separation, that obtained the sphere with just the central crest of
the wavy surface, having a reduced equivalent radius
\begin{equation}
\frac{1}{R_{eff}}=\frac{1}{R}+4\pi^{2}\frac{A}{\lambda^{2}}%
\end{equation}
If we consider the pull-off value with this reduced radius, this leads to
$\quad$
\begin{equation}
\widehat{P}_{\min,large\beta_{G}}=-\frac{R_{eff}}{R}=-\frac{1}{1+4\pi
^{2}\alpha_{G}}=-\frac{\pi^{3/2}\alpha_{KLJ}\sqrt{\beta_{G}}}{\pi^{3/2}%
\alpha_{KLJ}\sqrt{\beta_{G}}+4\pi^{2}}%
\end{equation}
which goes to 1 both for large $\beta_{G}$ and large $\alpha_{KLJ}\ $\ as it
is clear from Fig.5.

Combinations of large $\beta_{G}$ and small $\alpha_{KLJ}\ $\ shows this
reduction is not too large. Guduru suggested there is an optimal wavelength
where the enhancement is highest; however, this must necessarily be just a
different rewording of the dependences on the parameters. At low $\beta_{G}$,
there is so far no real limit on the enhancement, if it wasn't for the
conditions on the following paragraphs.

\begin{center}
$%
\begin{array}
[c]{cc}%
{\includegraphics[
height=2.4526in,
width=3.659in
]%
{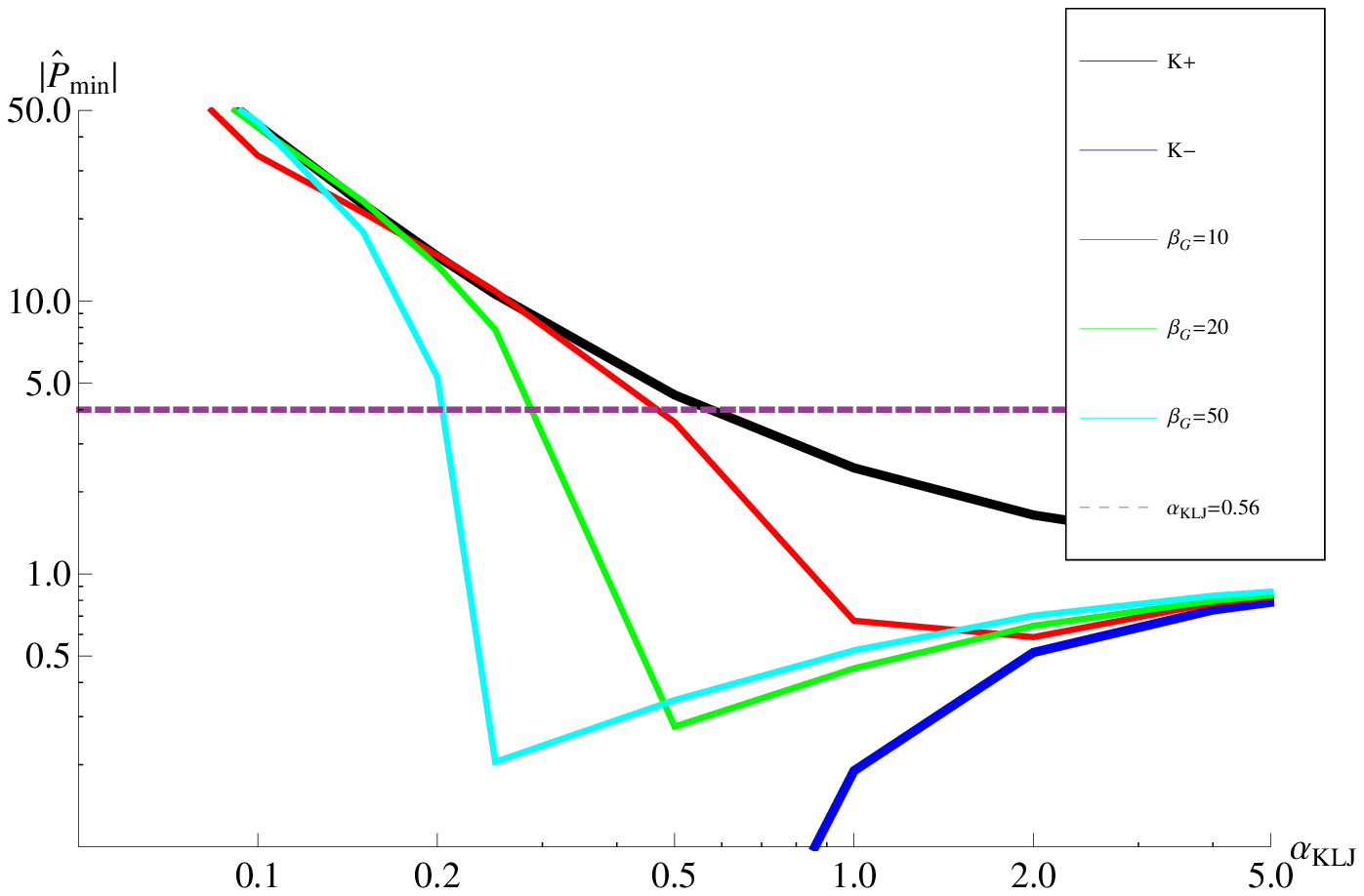}%
}
&
\end{array}
$

Fig.5 - Enhancement of pull-off. Black thick line is obtained with Kesari
equation on unloading, blue line with Kesari equation on loading, and
red,green,cyan are obtained with the full Guduru solution with $\beta
_{G}=10,20,50$
\end{center}

\subsection{Contradiction of the simply connected area}

The enhancement predicted from the Kesari envelope curve on unloading becomes
arbitrarily large at low $\alpha_{KLJ}$ and indeed also Guduru (2007) in
Fig.11, shows values of the order of 30-40 factor of increase, with values
higher than 15 being actually measured in his experimental validation
paper\ (Guduru \&\ Bull, 2007). Guduru (2007) has a very preliminary
discussion about the validity of the assumption of simply connected contact
area, as for the case of non-adhesive contact, it can be cast very clearly in
terms of the parameter $\alpha_{G}=\frac{AR}{\lambda^{2}}$. In particular, to
have the gap function monotonically increasing, it is enough to have%
\begin{equation}
\alpha_{G}=\frac{AR}{\lambda^{2}}=\frac{1}{\pi^{3/2}\alpha_{KLJ}\sqrt
{\beta_{G}}}<\frac{1}{8.5761}=0.12
\end{equation}
or%
\begin{equation}
\beta_{G}>\beta_{\lim}=\left(  \frac{1.54}{\alpha_{KLJ}}\right)  ^{2}%
\end{equation}

Hence, at low $\alpha_{KLJ}$ enhancement in the Guduru-Kesari model can only
hold if $\beta_{G}$ is large, as otherwise separation in the contact may well
occur. As we are interested in the range $\alpha_{KLJ}<0.56$, the restriction
on the shape requires $\beta_{G}>7.6$ or larger: however, this conflicts with
the approximation in Kesari's envelope, so that in fact the enhancement should
be computed from the full Guduru's equation, and the actual minimum of the
curve is now much less than what expected from the envelope. An estimate of
this is seen from Fig.5 of Waters \textit{et al} (2009), where it is clear
that for $\alpha_{G}<0.12$ we are generally below $\left\vert \widehat
{P}_{\min}\right\vert =4.$ This is because at large $\beta_{G}$, we are
looking at the problem of a single asperity detaching, as we have seen in
point 1) of the discussion above, which reduced adhesion, instead of enhancing it.

In Fig.6, we plot the boundary defined by $\beta_{G}=\beta_{\lim}$ with a red
curve: notice that this corresponds to increasingly high values of $\beta_{G}$
the more we move towards the left. Only points below this curve are
"certainly" satisfying the monotonicity of the punch and hence of the simply
connected contact. It is clear that within this region, the enhancement is
lower than 4 which is also the highest amplification with Johnson parameter
$\alpha_{KLJ}>0.56$.

\begin{center}
$%
\begin{array}
[c]{cc}%
{\includegraphics[
height=2.1439in,
width=3.659in
]%
{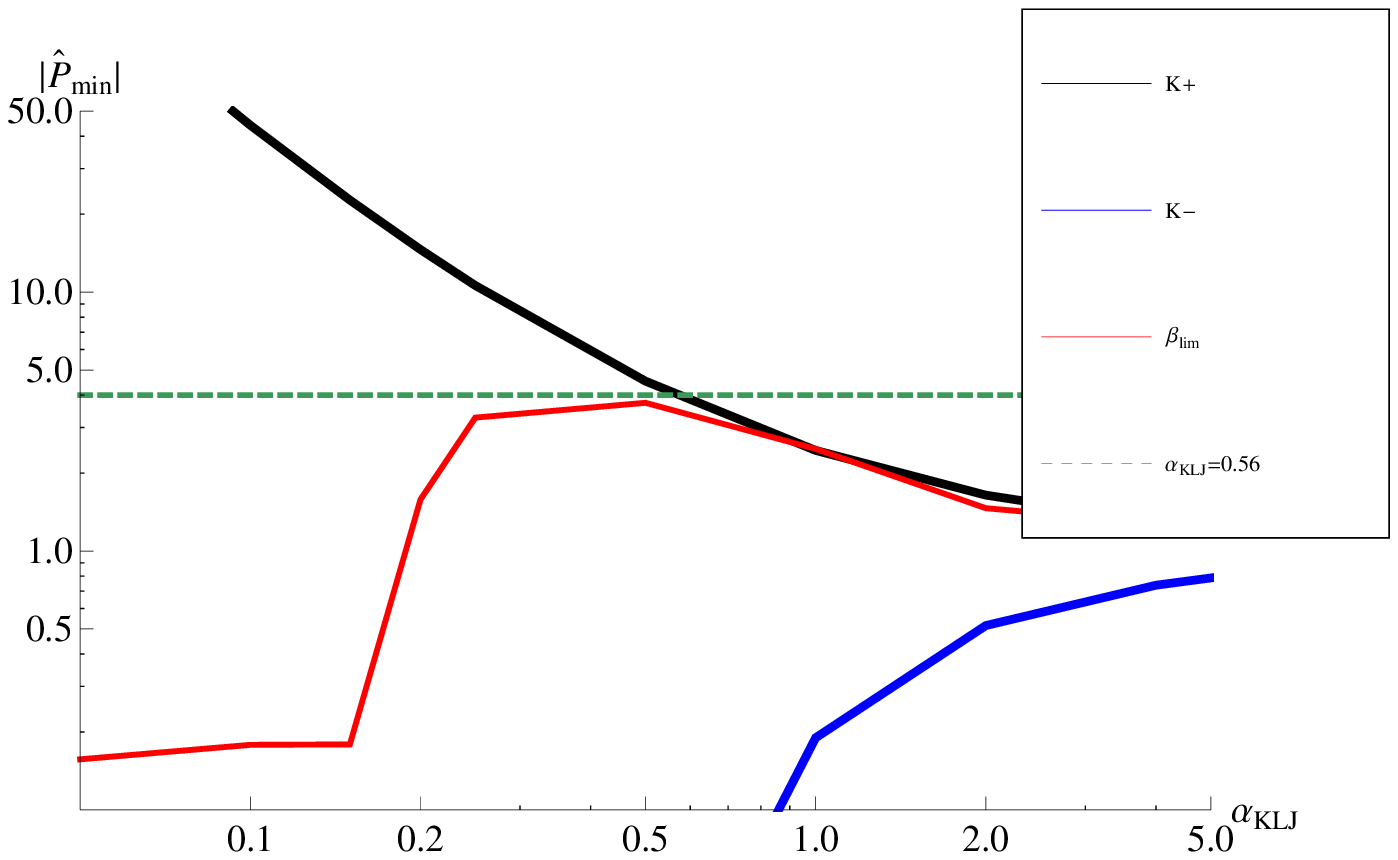}%
}
&
\end{array}
$

Fig.6 - The boundary of enhancement of pull-off. Black thick line is obtained
with Kesari equation on unloading, in blue curve Kesari equation on loading,
and red curve is $\beta_{G}=\beta_{\lim}\left(  \alpha_{KLJ}\right)  $
\end{center}

Guduru (2007) notices that the condition $\alpha_{G}<0.12$ is too restrictive
on two grounds: one, because of the effect of adhesion, which can only make
the likelihood of contact greater: indeed, even from the single sinusoid
solution that we described in the first paragraph, we know that the pressure
to reach full contact decreases from the purely mechanical one $p^{\ast}=\pi
E^{\ast}A/\lambda$ and could be defined as $p^{\ast}\left(  \alpha
_{KLJ}\right)  $, see Afferrante \textit{et al} (2015). This is important as
we know from that analysis, that once in full contact, there is little chance
to return to separated contact (Johnson (1995) has to postulate a flaw at the
interface, or else there is theoretical strength as the only possible limit),
other than from the contact edge. Second, because the monotonicity of the
punch profile is reobtained for large compressions, i.e. loading the contact
substantially before starting the unloading. In particular, from the function
of the profile, the condition is immediately found as
\begin{equation}
\frac{1}{4\pi^{2}\alpha_{G}}>-\frac{\sin\left(  2\pi r/\lambda\right)
}{\left(  2\pi r/\lambda\right)  }%
\end{equation}
which has a maximum giving the absolute $\alpha_{G}<0.1166$, but otherwise
should be satisfied for $r>r_{c}$, where one choice is
\[
\frac{r_{c}}{\lambda}=2\pi A\frac{R}{\lambda^{2}}=2\pi\alpha_{G}>>2\pi
\cdot0.12=0.75
\]
and indeed $1>-\sin\left(  4\pi^{2}A\frac{R}{\lambda^{2}}\right)  $. This is
not further discussed in Guduru (2007), but here we will make more
considerations. First of all, we should notice that to have the radius of
contact less than, say, 1/4 of the radius of the sphere, we need $2\pi\frac
{A}{\lambda}R<\frac{R}{4}$, which gives
\begin{equation}
\frac{A}{\lambda}<\frac{1}{8\pi}=0.04
\end{equation}
This together with $\alpha_{G}=\frac{AR}{\lambda^{2}}>0.12$ gives $\frac
{R}{\lambda}>\frac{0.12}{A/\lambda}=3$ or higher. With respect to
$\alpha_{KLJ}$, the condition on $\frac{A}{\lambda}<\frac{1}{8\pi}$ leads to
$\alpha_{KLJ}>4\sqrt{\frac{l_{a}}{\pi A}}$, where $l_{a}=w/E^{\ast}.$ Since we
want $\alpha_{KLJ}<0.56$, for example $\alpha_{KLJ}=0.1$, then we need
$\frac{l_{a}}{A}<\pi\left(  \frac{0.1}{4}\right)  ^{2}$ or $A>\allowbreak
510l_{a}$. But $A<\frac{\lambda}{8\pi}$ so that $\lambda>12818l_{a},$ and
therefore $R>38\,454l_{a}$, so we need big enough spheres with respect to
adhesion characteristic length.

Let us further estimate the pressure distribution (from Guduru (2007), under
adhesionless contact only) with respect to $p^{\ast}=\pi E^{\ast}A/\lambda,$
noting that with $\frac{A}{\lambda}=0.04$ this corresponds to a quite high
compressive stress, just 1/10 of the elastic modulus, with the warning that
finite strains and other deviations may occur. We could convince ourselves
that the waviness has been squashed out mechanically if the local pressure is
positive. Now, rearranging Guduru's equation in the form%
\begin{equation}
\frac{p}{p^{\ast}}\left(  \frac{r}{a}\right)  =2\left(  \frac{1}{\pi^{2}%
\alpha_{G}}+2\right)  \frac{a}{\lambda}\sqrt{1-\left(  \frac{r}{a}\right)
^{2}}+\int_{r/a}^{1}\frac{H_{0}\left(  2\pi\frac{x}{a}\frac{a}{\lambda
}\right)  }{\sqrt{\left(  \frac{x}{a}\right)  ^{2}-\left(  \frac{r}{a}\right)
^{2}}}d\frac{x}{a}-2\pi\left(  \frac{a}{\lambda}\right)  \int_{r/a}^{1}%
\frac{\frac{x}{a}H_{1}\left(  2\pi\frac{x}{a}\frac{a}{\lambda}\right)  }%
{\sqrt{\left(  \frac{x}{a}\right)  ^{2}-\left(  \frac{r}{a}\right)  ^{2}}%
}d\frac{x}{a}%
\end{equation}
we can plot it for $\frac{a}{\lambda}=\frac{r_{c}}{\lambda}=2\pi\alpha_{G}$
(Fig.7a) and $\alpha_{G}=0.05,0.1,1,5$, as well as for even higher contact
radius $\frac{a}{\lambda}=3\frac{r_{c}}{\lambda}=6\pi\alpha_{G}$.(Fig.7b). The
results show that at the very low $\alpha_{G}=0.075$ (thick black line) the
pressure is indeed always compressive and this should occur for any value of
the contact area. However, some tension appears for $\alpha_{G}=0.15$ (thick
blue line) or higher values, which is where the "local" monotonicity condition
postulated by Guduru (2007) should also suggest compression always for any
value of the contact area. Only with triple radius of contact, as in\ Fig.7b,
we do find always compressive stresses. In the case of adhesion, since we are
interested in the case $\alpha_{KLJ}<0.56 $, there is no guarantee therefore
that the "local" condition of Guduru (2007) justifies the simple connected
area assumption. This, together with the fact that the pressure may be
impractically large, suggests that we may not be able to observe the actual
enhancement predicted by the simply connected contact area solution.

\begin{center}
$%
\begin{array}
[c]{cc}%
{\includegraphics[
height=2.4241in,
width=3.659in
]%
{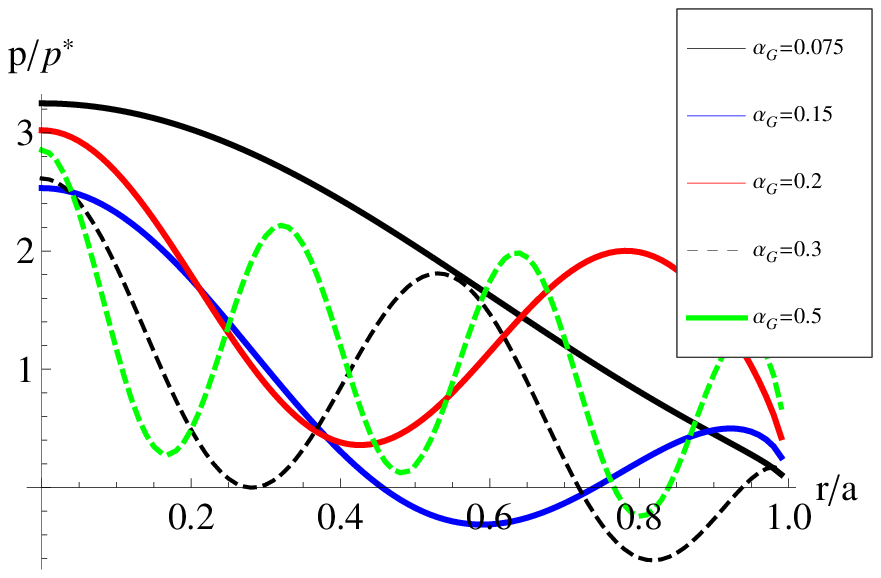}%
}
& (a)\\%
{\includegraphics[
height=2.4241in,
width=3.659in
]%
{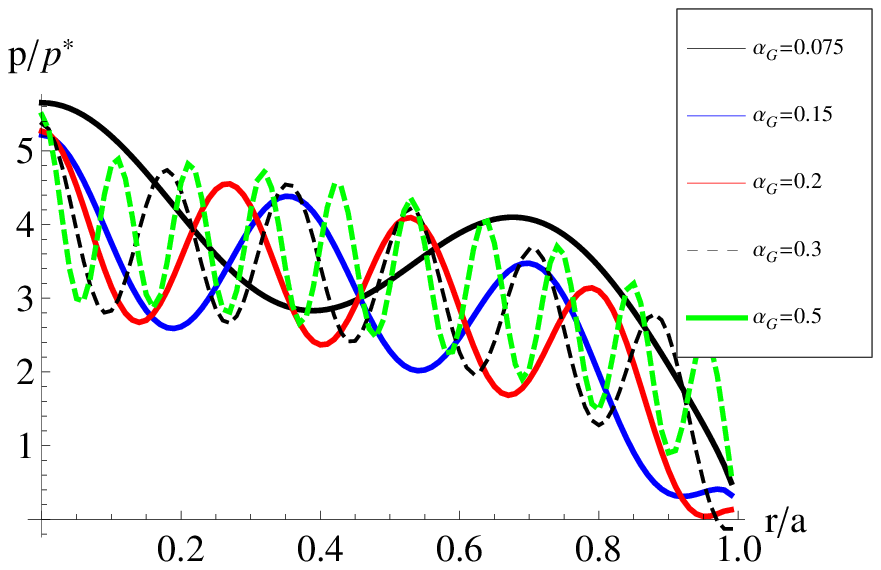}%
}
& (b)
\end{array}
$

Fig.7 - Pressure distributions with pure mechanical contact, for $\alpha
_{G}=0.075,0.15,0.2,0.3,0.5$ and for contact radius (a) $\frac{a}{\lambda
}=\frac{r_{c}}{\lambda}=2\pi\alpha_{G}$ or (b) $\frac{a}{\lambda}=3\frac
{r_{c}}{\lambda}=3\times2\pi\alpha_{G}$
\end{center}

\section{Deviation from the JKR regime}

The transition in adhesion from Bradley-rigid behaviour\footnote{It is more
precise to say from rigid behaviour to JKR regime, as the DMT solution,
commonly referred to as the limit for low Tabor parameter, has indeed various
forms, and most of them not exact.} to fully elastic JKR behaviour is well
known for the sphere. It depends on the well known Tabor parameter
\begin{equation}
\mu=\frac{\sigma_{th}}{E^{\ast}}\left(  \frac{9RE^{\ast}}{2\pi w}\right)
^{1/3}%
\end{equation}
where $\sigma_{th}$ is the theoretical strength and we have defined the
parameter with reference to the sphere, but we could also define a value
appropriate for the roughness.

Anyway, the phenomenon of enhancement of toughness and strength because of
surface waviness has been shown to be restricted primarily to the JKR adhesion
regime in Waters \textit{et al} (2009), or $\mu>1$. This is important in view
of application of the Guduru model to rough surfaces. Guduru and Kesari have
considered very soft materials with very small amounts of roughness. If we
consider the more standard situation of macroscopic contacts with many
asperities, the local behaviour of asperities will not show any enhancement.
However, in intermediate situations the situation is unclear.

However, a further remark is that the limit of "rigid" roughness is not
necessarily that of the sphere without roughness, as it seems suggested in
Waters \textit{et al} (2009). Indeed, first of all, we have the limit at large
$\beta_{G}$ which should reduce the pull-off to the case of contact between
the sphere and the first crest of the waviness. Secondly, we have at our
disposal a solution for rigid adhesion between a spherical particle and a
rough plane (simplified with a single small asperity, and otherwise smooth
plane), due to Rumpf (1990), later modified by Rabinovich et al
(2000)\footnote{Rabinovicz et al (2000) modified this to take into account a
one-scale roughness but this simply changes the radius of the asperity into
$r=1.485rms $, but the behaviour is the same. In the nanoscale roughness
regime, a dramatic decrease in adhesion force is predicted for this size
adhering particle.} which is obtained applying Derjaguin's approximation and
contains two terms: a first term represents the interaction sphere/asperity
(which increases with $r$), and \textquotedblleft noncontact\textquotedblright%
\ particle/flat separated by the height (radius) of the asperity (decreasing
with $r$)%
\begin{equation}
\frac{P_{ad}}{P_{smooth}}=\frac{1}{1+R/r}+\frac{1}{\left(  1+r/H_{0}\right)
^{2}}%
\end{equation}
where $H_{0}$ is some atomic size length scale. This equation has been used
very much in the area of particle adhesion and powder technology, as well as
drug delivery, semiconductor fabrication, xerographic processes, and paint
formulation or aerosol formation, amongst others) and shown to be reasonably
in agreement with experiments particularly at nanoscale -- despite its
simplicity and the strong assumption of negligible elastic deformations.

This Rumpf-Rabinovich model is very useful for our purpose as, for both
extremely small "roughness" (subatomic roughness) $r/R\rightarrow0$, or very
large radius of asperity $r$ (when the contact is essentially between the
sphere and a single asperity which in this case has grown in curvature so that
it is a flat plane itself), this model leads to the value for the particle
alone on the flat surface $P_{smooth}$ which is in this case the Bradley
result for the sphere, $2\pi Rw$. This is the value Waters \textit{et al}
(2009) seem to obtain for $\mu\rightarrow0.$The model shows therefore with
increasing $r/R$ first a decrease of adhesion due to roughness and then, after
reaching a minimum, an increase towards a linear trend in R of the Bradley equation.

\bigskip In Guduru's notation $r=\frac{\lambda^{2}}{A\left(  2\pi\right)
^{2}}$ and
\begin{equation}
\frac{r}{R}=\frac{1}{\left(  2\pi\right)  ^{2}\alpha_{G}}%
\end{equation}
and therefore the Rumpf-Rabinovich model gives%
\begin{equation}
\frac{P_{ad}}{P_{smooth}}=\frac{1}{1+\left(  2\pi\right)  ^{2}\alpha_{G}%
}+\frac{1}{\left(  1+\frac{\lambda^{2}}{AH_{0}\left(  2\pi\right)  ^{2}%
}\right)  ^{2}}%
\end{equation}

This formula has been shown to work pretty well with nano and atomic size
roughness, with hard solids. In our case, as we are uncertain especially of
the case $\alpha_{G}>0.12$, corresponding to $\frac{r}{R}<0.21$, it is clear
that the first term is small (interaction sphere asperity), of $0.174$ and
less. On the other hand, the second term giving the interaction with the
smooth plane separated by the hemiasperity is also extremely small too when
$r>>H_{0}$ (atomic size) as it is common. Suppose $R=10^{9}H_{0}$, and\ the
Rumpf-Rabinovich model gives
\begin{equation}
\frac{P_{ad}}{P_{smooth}}=\frac{1}{1+\left(  2\pi\right)  ^{2}\alpha_{G}%
}+\frac{1}{\left(  1+\frac{10^{9}}{\left(  2\pi\right)  ^{2}\alpha_{G}%
}\right)  ^{2}}%
\end{equation}

We explore the range $r/H_{0}=1,10^{9}$, i.e. $\alpha_{G}=\frac{1}{\left(
2\pi\right)  ^{2}},\frac{10^{9}}{\left(  2\pi\right)  ^{2}}$ in Fig.8 below,
which clearly shows that reduction of pull-off of various orders of magnitude
for $\alpha_{G}$ increasing in the range where we cannot assume full contact.

\begin{center}
$%
\begin{array}
[c]{cc}%
{\includegraphics[
height=2.2857in,
width=3.659in
]%
{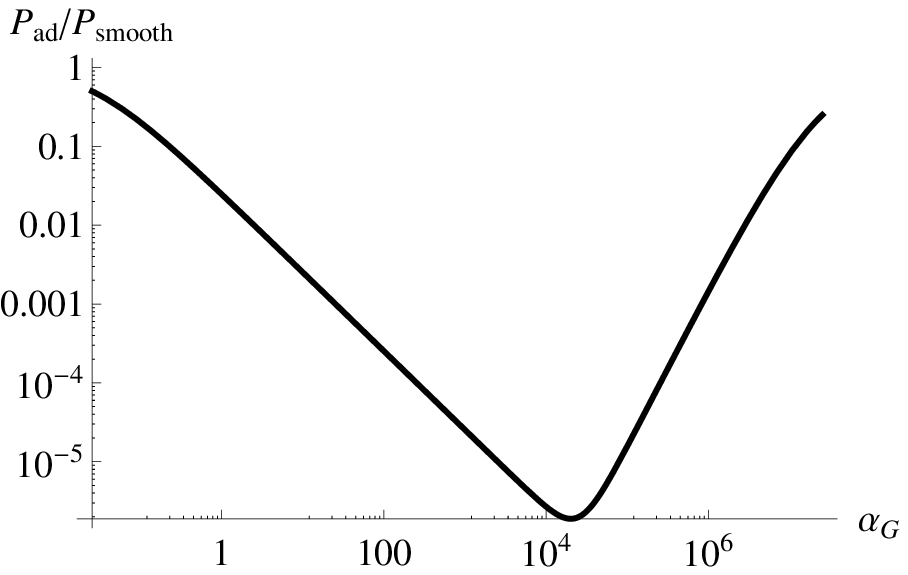}%
}
& (a)
\end{array}
$

Fig.8 - The Rumpf-Rabinovich model as applied to the Guduru problem
\end{center}

\bigskip It is clear that, by assuming a simply connected contact area, Waters
\textit{et al} (2009) found only a very limited range of possible decay of
adhesion, although they should have found at least the regime at high
$\beta_{G}$ where the contact is expected between the sphere and the first
crest of the sinusoidal waviness, therefore reduced with respect to the
Bradley result for the sphere.

\section{Conclusions}

We have revisited the Guduru model, and discussed the possible reasons for the
limitation of the very high enhancement of adhesion found in that model. In
particular, we have observed a reduced dependence on the parameters, we have
identified the Johnson parameter for single sinusoidal contact to govern also
the amplification of adhesion, at least in an asymptotic regime where the
envelope solution by Kesari et al holds.

Finally, we have shown that, as very large amplification is expected from the
Guduru model, even greater reduction is expected in the separated contact
regime, as estimated from a rigid model adhesion equation by Rumpf and
Rabinovich. The latter gives perhaps the lower bound of the adhesion, as
Guduru gives the upper bound.

\section{References}

Afferrante, L., Ciavarella, M., \& Demelio, G. (2015). Adhesive contact of the
Weierstrass profile. In Proc. R. Soc. A (Vol. 471, No. 2182, p. 20150248). The
Royal Society.

Briggs G A D and Briscoe B J (1977) The effect of surface topography on the
adhesion of elastic solids J. Phys. D: Appl. Phys. 10 2453--2466

Ciavarella, M., Demelio, G., Barber, J. R., \& Jang, Y. H. (2000). Linear
elastic contact of the Weierstrass profile. In Proceedings of the Royal
Society of London A: Mathematical, Physical and Engineering Sciences (Vol.
456, No. 1994, pp. 387-405). The Royal Society.

Ciavarella, M. (2016). Adhesive rough contacts near complete contact. in
press, Int J Mech Sci., arXiv preprint arXiv:1504.08240.

Fuller, K. N. G., \& Tabor, D. (1975). The effect of surface roughness on the
adhesion of elastic solids. Proc Roy Soc London A: 345, No. 1642, 327-342

Fuller, K.N.G. , Roberts A.D. (1981). Rubber rolling on rough surfaces J.
Phys. D Appl. Phys., 14, pp. 221--239

Gao, H., \& Yao, H. (2004). Shape insensitive optimal adhesion of nanoscale
fibrillar structures. Proceedings of the National Academy of Sciences of the
United States of America, 101(21), 7851-7856.

Greenwood, J. A. (2007). On the DMT theory. Tribology Letters, 26(3), 203-211.

\bigskip Guduru, P.R. (2007). Detachment of a rigid solid from an elastic wavy
surface: theory J. Mech. Phys. Solids, 55, 473--488

Guduru, P.R. , Bull, C. (2007). Detachment of a rigid solid from an elastic
wavy surface: experiments J. Mech. Phys. Solids, 55, 473--488

Hui, C. Y., Glassmaker, N. J., Tang, T., \& Jagota, A. (2004). Design of
biomimetic fibrillar interfaces: 2. Mechanics of enhanced adhesion. Journal of
The Royal Society Interface, 1(1), 35-48.

Johnson, K. L., K. Kendall, and A. D. Roberts. (1971). Surface energy and the
contact of elastic solids. Proc Royal Soc London A: 324. 1558.

Kesari, H., Doll, J. C., Pruitt, B. L., Cai, W., \& Lew, A. J. (2010). Role of
surface roughness in hysteresis during adhesive elastic contact. Philosophical
Magazine \& Philosophical Magazine Letters, 90(12), 891-902.

Kesari, H., \& Lew, A. J. (2011). Effective macroscopic adhesive contact
behavior induced by small surface roughness. Journal of the Mechanics and
Physics of Solids, 59(12), 2488-2510.

Kamperman, M., Kroner, E., del Campo, A., McMeeking, R. M., \& Arzt, E.
(2010). Functional adhesive surfaces with \textquotedblleft
gecko\textquotedblright\ effect: The concept of contact splitting. Advanced
Engineering Materials, 12(5), 335-348.

Persson, B.N.J. (2002). Adhesion between an elastic body and a randomly rough
hard surface, Eur. Phys. J. E 8, 385--401

Rabinovich, Y. I., Adler, J. J., Ata, A., Singh, R. K., \& Moudgil, B. M.
(2000). Adhesion between nanoscale rough surfaces: I. Role of asperity
geometry. Journal of Colloid and Interface Science, 232(1), 10-16.

Rumpf, H. Particle Technology, Chapman \& Hall, London/New York (1990)

Yao, H., \& Gao, H. (2006). Mechanics of robust and releasable adhesion in
biology: Bottom--up designed hierarchical structures of gecko. Journal of the
Mechanics and Physics of Solids, 54(6), 1120-1146.

Waters, J.F. Leeb, S. Guduru, P.R. (2009). Mechanics of axisymmetric wavy
surface adhesion: JKR--DMT transition solution, Int J of Solids and Struct 46
5, 1033--1042

\end{document}